\documentclass[12pt]{article}
\usepackage{epsfig}
\usepackage{wrapfig}
\usepackage{subfigure}
\usepackage{amsmath}
\usepackage{hhline}
\usepackage{amssymb}
\usepackage{times}
\usepackage{cite}
\usepackage[]{lineno}
\newlength{\dinwidth}
\newlength{\dinmargin}
\begin{document}  
%%%%%%%%%%%%%%%% Pre-defined commands, you can use for the most obvious notations
\newcommand{\pom}{{I\!\!P}}
\newcommand{\reg}{{I\!\!R}}
\newcommand{\slowpi}{\pi_{\mathit{slow}}}
\newcommand{\fiidiii}{F_2^{D(3)}}
\newcommand{\fiidiiiarg}{\fiidiii\,(\beta,\,Q^2,\,x)}
\newcommand{\n}{1.19\pm 0.06 (stat.) \pm0.07 (syst.)}
\newcommand{\nz}{1.30\pm 0.08 (stat.)^{+0.08}_{-0.14} (syst.)}
\newcommand{\fiidiiiful}{F_2^{D(4)}\,(\beta,\,Q^2,\,x,\,t)}
\newcommand{\fiipom}{\tilde F_2^D}
\newcommand{\ALPHA}{1.10\pm0.03 (stat.) \pm0.04 (syst.)}
\newcommand{\ALPHAZ}{1.15\pm0.04 (stat.)^{+0.04}_{-0.07} (syst.)}
\newcommand{\fiipomarg}{\fiipom\,(\beta,\,Q^2)}
\newcommand{\pomflux}{f_{\pom / p}}
\newcommand{\nxpom}{1.19\pm 0.06 (stat.) \pm0.07 (syst.)}
\newcommand {\gapprox}
   {\raisebox{-0.7ex}{$\stackrel {\textstyle>}{\sim}$}}
\newcommand {\lapprox}
   {\raisebox{-0.7ex}{$\stackrel {\textstyle<}{\sim}$}}
\def\gsim{\,\lower.25ex\hbox{$\scriptstyle\sim$}\kern-1.30ex%
\raise 0.55ex\hbox{$\scriptstyle >$}\,}
\def\lsim{\,\lower.25ex\hbox{$\scriptstyle\sim$}\kern-1.30ex%
\raise 0.55ex\hbox{$\scriptstyle <$}\,}
\newcommand{\pomfluxarg}{f_{\pom / p}\,(x_\pom)}
\newcommand{\dsf}{\mbox{$F_2^{D(3)}$}}
\newcommand{\dsfva}{\mbox{$F_2^{D(3)}(\beta,Q^2,x_{I\!\!P})$}}
\newcommand{\dsfvb}{\mbox{$F_2^{D(3)}(\beta,Q^2,x)$}}
\newcommand{\dsfpom}{$F_2^{I\!\!P}$}
\newcommand{\gap}{\stackrel{>}{\sim}}
\newcommand{\lap}{\stackrel{<}{\sim}}
\newcommand{\fem}{$F_2^{em}$}
\newcommand{\tsnmp}{$\tilde{\sigma}_{NC}(e^{\mp})$}
\newcommand{\tsnm}{$\tilde{\sigma}_{NC}(e^-)$}
\newcommand{\tsnp}{$\tilde{\sigma}_{NC}(e^+)$}
\newcommand{\st}{$\star$}
\newcommand{\sst}{$\star \star$}
\newcommand{\ssst}{$\star \star \star$}
\newcommand{\sssst}{$\star \star \star \star$}
\newcommand{\tw}{\theta_W}
\newcommand{\sw}{\sin{\theta_W}}
\newcommand{\cw}{\cos{\theta_W}}
\newcommand{\sww}{\sin^2{\theta_W}}
\newcommand{\cww}{\cos^2{\theta_W}}
\newcommand{\trm}{m_{\perp}}
\newcommand{\trp}{p_{\perp}}
\newcommand{\trmm}{m_{\perp}^2}
\newcommand{\trpp}{p_{\perp}^2}
\newcommand{\alp}{\alpha_s}
\newcommand{\etamax}{\eta_{\rm max}}

\newcommand{\alps}{\alpha_s}
\newcommand{\sqrts}{$\sqrt{s}$}
\newcommand{\LO}{$O(\alpha_s^0)$}
\newcommand{\Oa}{$O(\alpha_s)$}
\newcommand{\Oaa}{$O(\alpha_s^2)$}
\newcommand{\PT}{p_{\perp}}
\newcommand{\JPSI}{J/\psi}
\newcommand{\sh}{\hat{s}}
\newcommand{\uh}{\hat{u}}
\newcommand{\MP}{m_{J/\psi}}
\newcommand{\PO}{I\!\!P}
\newcommand{\xbj}{x}
\newcommand{\xpom}{x_{\PO}}
\newcommand{\ttbs}{\char'134}
\newcommand{\xpomlo}{3\times10^{-4}}  
\newcommand{\xpomup}{0.05}  
\newcommand{\dgr}{^\circ}
\newcommand{\pbarnt}{\,\mbox{{\rm pb$^{-1}$}}}
\newcommand{\gev}{\,\mbox{GeV}}
\newcommand{\WBoson}{\mbox{$W$}}
\newcommand{\fbarn}{\,\mbox{{\rm fb}}}
\newcommand{\fbarnt}{\,\mbox{{\rm fb$^{-1}$}}}
\newcommand{\dsdx}[1]{$d\sigma\!/\!d #1\,$}
\newcommand{\eV}{\mbox{e\hspace{-0.08em}V}}
%
% Some useful tex commands
%
\newcommand{\qsq}{\ensuremath{Q^2} }
\newcommand{\gevsq}{\ensuremath{\mathrm{GeV}^2} }
\newcommand{\et}{\ensuremath{E_t^*} }
\newcommand{\rap}{\ensuremath{\eta^*} }
\newcommand{\gp}{\ensuremath{\gamma^*}p }
\newcommand{\dsiget}{\ensuremath{{\rm d}\sigma_{ep}/{\rm d}E_t^*} }
\newcommand{\dsigrap}{\ensuremath{{\rm d}\sigma_{ep}/{\rm d}\eta^*} }

%%% Dstar stuff
\newcommand{\dstar}{\ensuremath{D^*}}
\newcommand{\dstarp}{\ensuremath{D^{*+}}}
\newcommand{\dstarm}{\ensuremath{D^{*-}}}
\newcommand{\dstarpm}{\ensuremath{D^{*\pm}}}
\newcommand{\zDs}{\ensuremath{z(\dstar )}}
\newcommand{\Wgp}{\ensuremath{W_{\gamma p}}}
\newcommand{\ptds}{\ensuremath{p_t(\dstar )}}
\newcommand{\etads}{\ensuremath{\eta(\dstar )}}
\newcommand{\ptj}{\ensuremath{p_t(\mbox{jet})}}
\newcommand{\ptjn}[1]{\ensuremath{p_t(\mbox{jet$_{#1}$})}}
\newcommand{\etaj}{\ensuremath{\eta(\mbox{jet})}}
\newcommand{\detadsj}{\ensuremath{\eta(\dstar )\, \mbox{-}\, \etaj}}

% Journal macro
\def\Journal#1#2#3#4{{#1} {\bf #2} (#3) #4}
\def\NCA{\em Nuovo Cimento}
\def\NIM{\em Nucl. Instrum. Methods}
\def\NIMA{{\em Nucl. Instrum. Methods} {\bf A}}
\def\NPB{{\em Nucl. Phys.}   {\bf B}}
\def\PLB{{\em Phys. Lett.}   {\bf B}}
\def\PRL{\em Phys. Rev. Lett.}
\def\PRD{{\em Phys. Rev.}    {\bf D}}
\def\ZPC{{\em Z. Phys.}      {\bf C}}
\def\EJC{{\em Eur. Phys. J.} {\bf C}}
\def\CPC{\em Comp. Phys. Commun.}

\begin{titlepage}

\noindent
%\begin{flushleft}
%{\tt DESY YY-NNN    \hfill    ISSN 0418-9833} \\
%{\tt Month YYYY}                  \\
%\end{flushleft}

\noindent
%Date:          [today  instruction is preferred] \\ %\today      \\
%Version:       Preparatives 0.1,0.2...; 1st draft: 1.0, 1.1...; 2nd Draft 2.0..., Final Reading 3.0,3.1...      \\
%Editors:            \\
%Referees:           \\
%Comments by         \\
\noindent
%%%%%%%%%%%%%%%%%%%%%%%%%%%%%%%%%%%%%%%%%%%%%%%%%%%%%%%%%%%%%%%%%%%%%%%%%%%%%%%%%%%%%%%%%%%%%%
%%%%%%%%%%%% For conference papers  %%%%%%%%%%%%%%%%%%%%%%%%%%%%%%%%%%%%%%%%%%%%%%%%%%%%%%%%%%
%%%%%%%%%%%% coment the header and fill the right conference
%%%%% {\it {\large version of \today}} \\[.3em]

\vspace{2cm}
\begin{center}
\begin{Large}

{\bf Returns in futures markets and $\nu=3$ t-distribution
%\\(version of \today )
}
\end{Large}

\vspace{2cm}

Laurent Schoeffel \\~\\
CEA Saclay, Irfu/SPP, 91191 Gif/Yvette Cedex, \\
France
%%\end{Large}
\end{center}

\vspace{2cm}

\begin{abstract}

The probability distribution of log-returns of
financial time series, sampled at high frequency, 
is the basis for any further developments in quantitative finance.
In this letter, we present experimental results based on a large set of
time series on futures.
Then, we show that the t-distribution with $\nu \simeq 3$ gives a nice description
of almost all data series.
This appears to be
a quite general result that stays robust on a large set of any financial data as well as on a wide range of sampling frequency
of these data,
below one hour.

\end{abstract}

\vspace{1.5cm}

\begin{center}
%%To be submitted to {\it The Journal of Investment Strategies}
\end{center}

\end{titlepage}

%          THE PAPER DRAFTS HAVE NO AUTHORLIST
%
%          FOR PAPER ISSUED FOR THE FINAL READING 
%          COPY THE AUTHOR AND INSTITUTE LISTS 
%          INTO YOUR AREA
%
% from /h1/iww/ipublications/h1auts.tex 
%          AND UNCOMMENT THE NEXT THREE LINES 
%
%\begin{flushleft}
%  \input{h1auts}
%\end{flushleft}
%
% Please not that the author list may need re-formatting.

%%\newpage
%=========================================================================
\section{Introduction}
%=========================================================================
%\noindent

Returns in financial time series are the most fundamental inputs to quantitative finance.
To a certain extend, they provide some insights in the the dynamical content of the market.
In this letter, we consider several financial series on futures, using always a five minutes
sampling. Each future contract is characterized by a price series $S(t)$, from which
we extract the log-returns $x(t)=\log [S(t+1)/S(t)]$.  The analysis is then driven on these
log-returns $x(t)$.

From standard quantitative analysis \cite{a1,a2,a3,a4}, we know that the distribution of log-returns $x(t)$,
namely $P(x)$, can be written quite generally as
\begin{equation}
\label{ini}
P(x)=\frac{1}{Z} \exp( - \frac{2}{D} w(x)/2)
\end{equation}
where $w(x)$ is an objective function and $Z$ a normalization factor. In particular, it can be shown easily that
 $w(x)$ can be derived by minimizing a generating functional $F[w(x)]$, subject to some constraints on
the mean value of the objective function. It reads
\begin{equation}
\label{ini2}
F = \int dx P(x) \left[ \log P(x) + w(x)/D -\lambda \right]
\end{equation}
where $\lambda$ is an arbitrary constant.

In addition, the expression given in Eq. (\ref{ini})  for the probability distribution 
can also be seen as the outcome of an equation of motion for $x(t)$. From Eq. (\ref{ini}) and (\ref{ini2}),
we can express the stochastic process $x(t)$ as 
 a Markovian  process of the form \cite{a1,a2,a3,a4}
\begin{equation}
\label{eq1}
\frac {dx}{dt} = f(x) + g(x) \epsilon(t)
\end{equation}
where $\epsilon(t)$ is a Gaussian process satisfying $<\epsilon(t)\epsilon(t')>=D \delta(t-t')$ and $<\epsilon(t)>=0$.
In Eq. (\ref{eq1}), functions $f$ and $g$ depends only on $x(t)$. Adopting the It\^o convention
\cite{a1,a2,a3,a4}, the distribution 
function $P(x,t)$, associated with this equation of motion (Eq. (\ref{eq1})), is given by the 
following Fokker-Planck equation
\begin{equation}
\label{eq2}
\frac{\partial P(x,t)}{\partial t}= 
\frac{\partial^2}{\partial x^2} [\frac{D}{2} g^2(x) P(x,t)]
-
\frac{\partial}{\partial x} [f(x) P(x,t)]
\end{equation}
From Eq. (\ref{eq2}), we can finally extract the stationary solution for $P(x)$ in the form of Eq. (\ref{ini})

\begin{equation}
\label{eq3}
P(x) = \frac{1}{Z} \exp \left[ -\frac{2}{D} \int dx \frac{D g \frac{dg}{dx} -f}{g^2}   \right]
\end{equation}

Whether Eq. (\ref{eq1}), (\ref{eq2}) and (\ref{eq3}) can be related to real data on financial markets is
not granted. Therefore, we need to compare predictions derived from these equations to real
data. As mentioned above, we use financial time series on different futures, using a five minutes sampling.

%=========================================================================
\section{Parameterizations of $P(x)$}
%=========================================================================
\label{param}

In Eq. (\ref{eq1}), (\ref{eq2}) and (\ref{eq3}), functions $f$ and $g$ are not specified and any choice can be considered.
Obviously, only specific choices will have a chance to get a reasonable agreement with real data.
For example, let us consider three cases:
\begin{itemize}

\item[(i)]
If $f(x)=-x$ and $g(x)=1$, we obtain $P(x)Z = \exp(-x^2/D)$, and thus we predict a Gaussian shape for the log-returns
distribution.
\item[(ii)]
In the more general case where $f(x)=\lambda g \frac{dg}{dx}$ and
$g$ is not constant, we obtain 
$$
P(x)Z = \frac{1}{g^{2(1-\lambda/D)}}
$$
and thus we predict non-Gaussian shape for the log-returns
distribution.
\item[(iii)] Let us specify the case (ii).
Introducing a constant $\nu$ and defining the two functions $f$ and $g$ as
$
f(x)=\frac{2x}{\nu} (1+\frac{x^2}{\nu})
$ and
$
g(x)=(1+\frac{x^2}{\nu})
$,
we get
\begin{equation}
\label{nu}
P(x)Z = \frac{1}{(1+\frac{x^2}{\nu})^{(\nu+1)/2}}.
\end{equation}
In this scenario $P(x)$ follows the so-called t-distribution.
It depends on one parameter $\nu$ to be fitted on real data, for normalized log-returns.
Let us notice that Eq. (\ref{nu}) is equivalent to the
q-exponential form of Ref. \cite{ts1,ts2}
\begin{equation}
\label{q}
P(x)Z = \frac{1}{(1+x^2 \frac{q-1}{3-q})^{1/(q-1)}}.
\end{equation}
where we have conserved the notations of Ref. \cite{ts1,ts2}.
Obviously, Eq. (\ref{nu}) and (\ref{q}) are directly related by
$(\nu+1) /2 = 1/(q-1)$. In particular, $\nu=3$ is equivalent to $q=1.5$.

\end{itemize}

%=========================================================================
\section{Experimental Analysis of $P(x)$}
%=========================================================================

\begin{figure}[htbp]
  \begin{center}
    \includegraphics[width=0.45\textwidth]{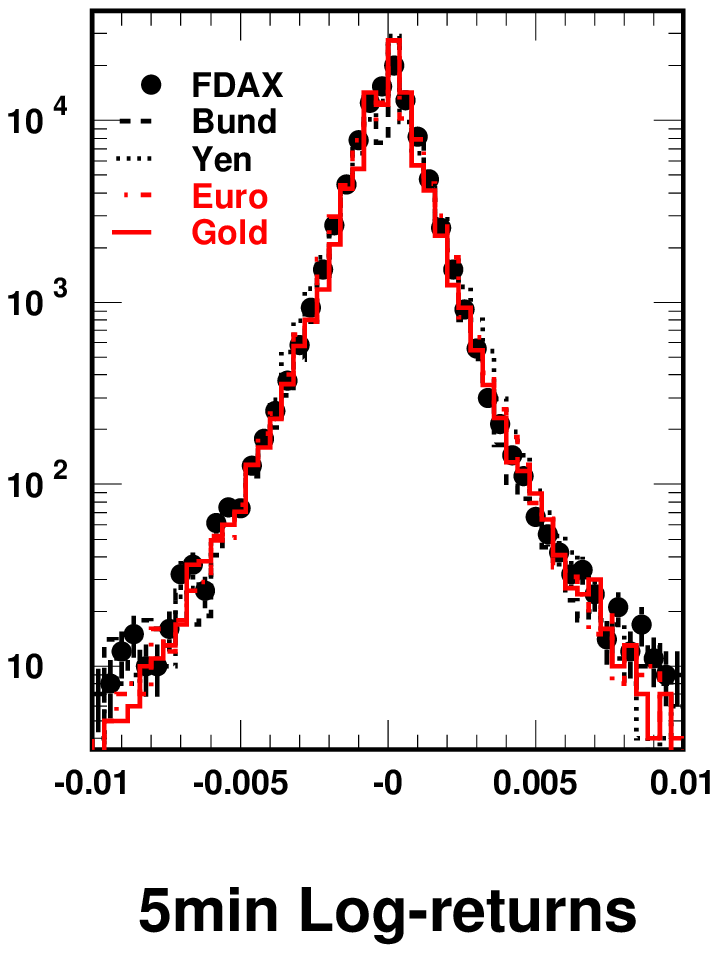}
    \includegraphics[width=0.45\textwidth]{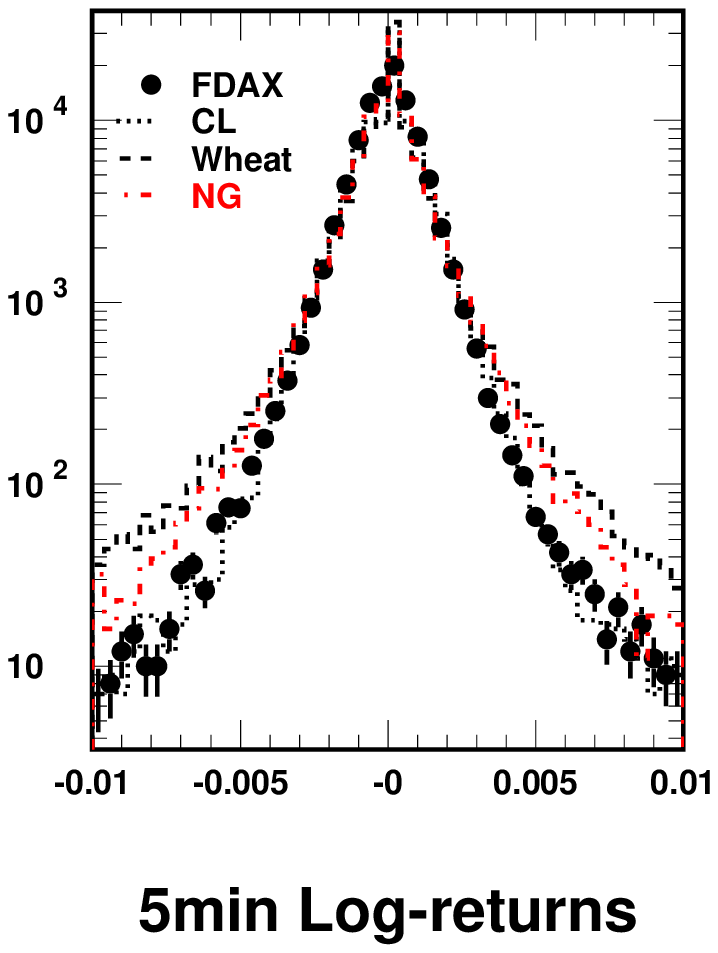}
  \end{center}
  \caption{Log-returns  $x$ (five minutes sampling) for a large set of futures.
To make the comparison, we have scaled $x$ for all futures to the volatility of the DAX future (FDAX).
Left:
we observe that futures on DAX, Bund, Yen, Euro, Gold present the same probability distribution
for $x$, hence $P(x)$ is universal for all these data series once the volatility is normalized to the same value.
Right: we observe that data series on CL (Crude Oil) follows the same $P(x)$ as FDAX, but
other data series on Wheat and NG (Natual Gaz) exhibit some larger tails.}
\label{fig1}
\end{figure}

In Fig. \ref{fig1}, we present the log-returns  $x$ (five minutes sampling) for a large set of futures.
To make the comparison, we have scaled $x$ for all futures to the volatility of the DAX future (FDAX).
On the left hand side (LHS) of Fig. \ref{fig1},
we observe that futures on DAX, Bund, Yen, Euro, Gold present the same probability distribution
for $x$, hence $P(x)$ is universal for all these data series once the volatility is normalized to the same value.
On the right hand side of Fig. \ref{fig1}, we provide comparisons with futures on commodities.
We observe that data series on CL (Crude Oil) follows the same $P(x)$ as FDAX, but
other data series on Wheat and NG (Natual Gaz) exhibit some larger tails.

\begin{figure}[htbp]
  \begin{center}
    \includegraphics[width=0.45\textwidth]{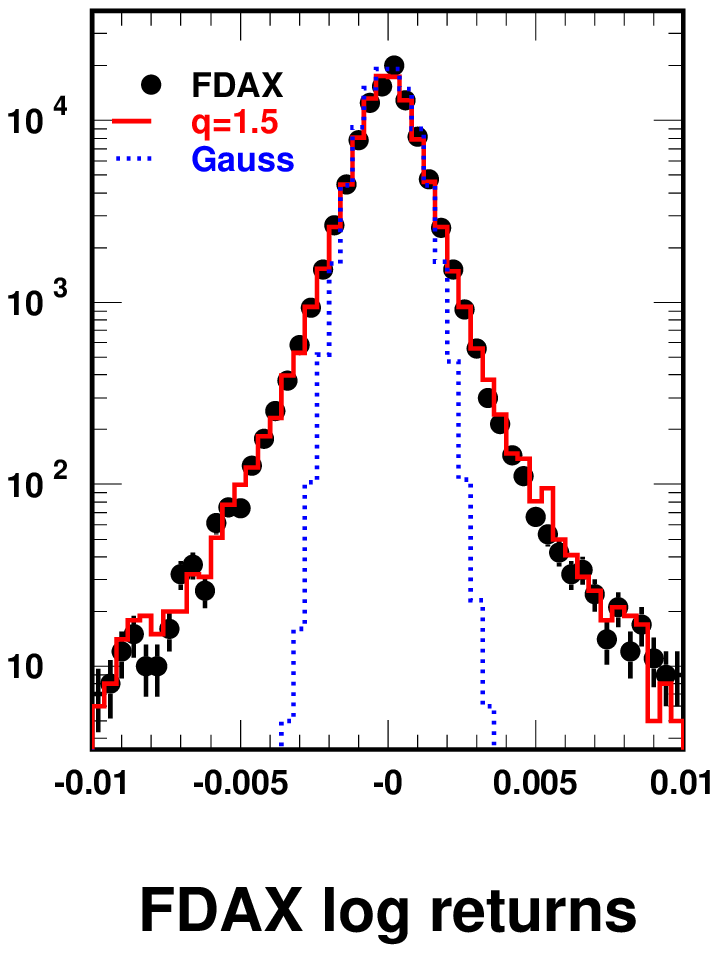}
    \includegraphics[width=0.45\textwidth]{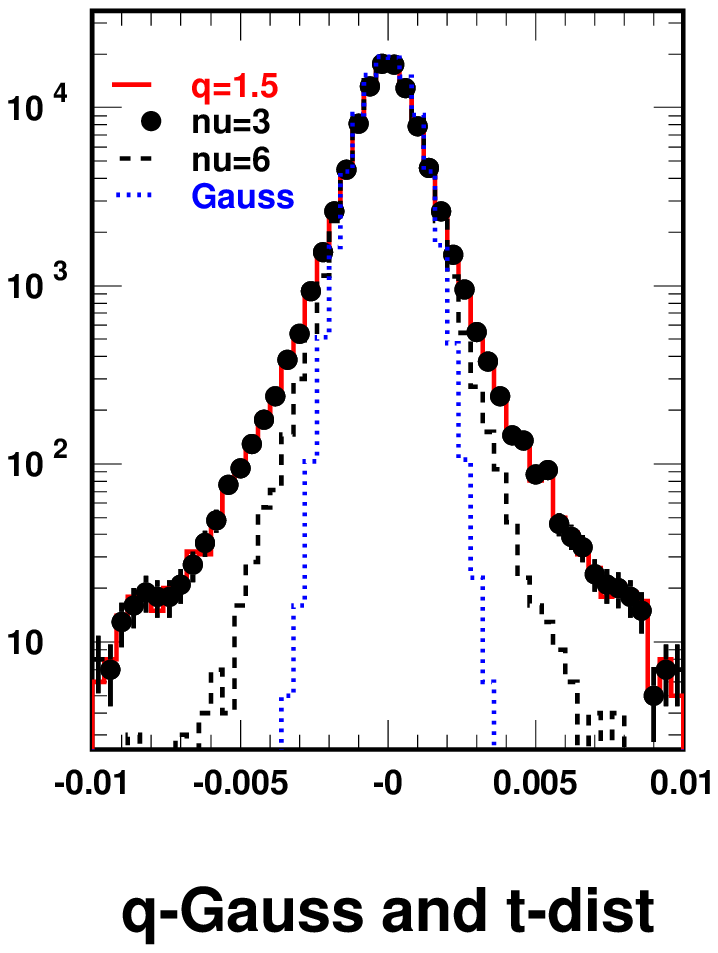}
  \end{center}
  \caption{
We show that the q-exponential probability
distribution  of Eq. (\ref{q}), with $q=1.5$, gives a good description of the data (Right).
It corresponds to $\nu=3$ in the form of the t-distribution of Eq. (\ref{nu}) (Left).
The Gaussian approximation fails to describe properly the data.
When $\nu$ is increased, we approach the Gaussian limit.
}
\label{fig2}
\end{figure}

We can use results developed in section \ref{param} in order to compare with experimental distributions $P(x)$
of Fig. \ref{fig1}. We use predictions exposed in cases (i) and (iii), respectively
the Gaussian and the q-exponential forms (or t-distribution).
Results are shown in Fig. \ref{fig2}.
For data, we only display $P(x)$ for FDAX.

In Fig. \ref{fig2}, we show that the q-exponential probability
distribution  of Eq. (\ref{q}), with $q=1.5$, gives a good description of the data.
Similarly, it corresponds to $\nu=3$ in the form of the t-distribution of Eq. (\ref{nu}).
Let us notice that futures with larger tails discussed above correspond to smaller values of $\nu$.
Also, we observe  in Fig. \ref{fig2} that the Gaussian approximation fails to describe properly the data.
On the right hand side of Fig. \ref{fig2}, we observe that when $\nu$ is increased above $3$ in
 Eq. (\ref{nu}), then $P(x)$ stands in the middle of the correct probability density and the Gaussian
approximation. When $\nu$ tends to infinity, the t-distribution recovers the Gaussian limit.

%=========================================================================
\section{Conclusion}
%=========================================================================

From the theory point of view, the probability distribution of log-returns of
financial time series, sampled at high frequency, can be expressed quite generally as $P(x)=\frac{1}{Z} \exp( - w(x)/D)$. In this
expression, the function $w(x)$ can be derived using the formalism of stochastic differential equations
applied to quantitative finance. Several cases have been considered in this letter. 
Then, we have shown that real data (with normalized log-returns) are compatible with a distribution of the form
$$
P(x) \propto \frac{1}{(1+\frac{x^2}{\nu})^{(\nu+1)/2}} \ \ \ , \ \ \ \nu \simeq 3.
$$
We have exemplified this property on a sample of data sets. However,
this is a quite general result that stays robust on a larger set of financial data as well as on a wide range of sampling frequency
of these data,
below one hour.
Interestingly also, this probability distribution corresponds to the equation of motion Eq. \ref{eq1}, with the
expressions of $f$ and $g$ given in case (iii). This can serve as a basis for a Monte-Carlo simulation of 
market data and subsequent risk analysis.

%=========================================================================

\newpage

\end{document}